\documentclass[prapplied,superscriptaddress,reprint,amsmath,amssymb,aps,longbibliography]{revtex4-2}

\usepackage{graphicx}
\usepackage{float}
\usepackage[
pdfstartview=FitH,
bookmarks=true,
breaklinks=true,
colorlinks=true,
linkcolor=blue,anchorcolor=blue,
citecolor=blue,filecolor=blue,
menucolor=blue,urlcolor=blue]{hyperref}

\begin{document}

\global\long\def\id{\mathbbm{1}}
\global\long\def\ui{\mathbbm{i}}
\global\long\def\ud{\mathrm{d}}

\title{Optimizing the image projection of spatially incoherent light from a multimode fiber}

\author{Ken Deng}
\affiliation{Department of Physics and State Key Laboratory of Low Dimensional Quantum Physics, Tsinghua University, Beijing, 100084, China}

\author{Zhongchi Zhang}
\affiliation{Department of Physics and State Key Laboratory of Low Dimensional Quantum Physics, Tsinghua University, Beijing, 100084, China}

\author{Huaichuan Wang}
\affiliation{Department of Physics and State Key Laboratory of Low Dimensional Quantum Physics, Tsinghua University, Beijing, 100084, China}

\author{Zihan Zhao}
\affiliation{Department of Physics and State Key Laboratory of Low Dimensional Quantum Physics, Tsinghua University, Beijing, 100084, China}

\author{Jiazhong Hu}
\email{hujiazhong01@ultracold.cn}
\affiliation{Department of Physics and State Key Laboratory of Low Dimensional Quantum Physics, Tsinghua University, Beijing, 100084, China}
\affiliation{Frontier Science Center for Quantum Information and Collaborative Innovation Center of Quantum Matter, Beijing, 100084, China}

\begin{abstract} 
We study the spatially incoherent light generated by a multimode fiber(MMF) in the application of image projection designed for the ultracold-atom experiments. Inspired by previous half-analytic methods concerning the incoherent light, here a full-numerical model is established to provide more quantitative descriptions, and part of results is compared with experiments. Particularly, our model about the MMF is also compatible with light propagation in free space. Based on this, we study both the intrinsic speckle and the perturbation robustness of a MMF light field, under the influence of light propagation and fiber parameters. We point out several guidelines about choosing the suitable MMF in creating a spatially incoherent light source, which is useful in the context of the ultracold-atom experiments associating with the optical potential projection. 
\end{abstract}

\maketitle

\section{Introduction}
Though lasers with both large temporal and spatial coherence continue to be a hot topic in various fields, light source with a reduced spatial coherence has also been gaining attentions these years. Spatially incoherent light has the natural property of resisting the external perturbations, and has been demonstrated potential applications in many fields, for example in free-space optical communications\cite{Shirai:03,Gbur:14} and light potential projections for ultracold atoms\cite{hudjf2o5h4eurcvau2dn88u0136bphoyt0rp16fi5w41k75n7r}. 

To reduce the spatial incoherence, 
one conventional approach is based on the time averaging, including both active and passive methods\cite{hudjf2o5h4eurcvau2dn88u0136bphoyt0rp16fi5w41k75n7r}.
However, at any particular time point, 
%both active and passive way are based on the time averaging, meaning that at any specific time point, 
the light field still maintains long-distance correlation. Thus the claim of an effective spatially incoherent light actually depends on the typical time scale of the targeted system. In other words, if the typical evolution frequency of the targeted system is faster than the time averaging frequency in producing spatial incoherence, this method will become invalid.

The targeted system that we are interested in is the ultracold atom gas, loaded into an optical dipole trap or an optical lattice \cite{WOS:000547609300001,WOS:000409455700037}. We intend to project arbitrary light potentials generated by a digital micromirror device (DMD) onto atoms to realize manipulations of quantum states \cite{DiCarli2024,10.1116/5.0026178,RevModPhys.94.041001,Zupancic2016UltrapreciseHB,Tajik2019DesigningAO}. In practice, unavoidable dusts, roughness, or reflections in the optical path will significantly distort the projected shapes and cause unwanted interference patterns if we use a coherent laser. In the past decade, this issue has been shown to be possibly solved by using a spatially incoherent light to illuminate the DMD chip \cite{Braun2024,ediss32357} instead. 

The time scale of our interested system is determined by the optical potentials depth, typically about several $10$ kHz \cite{Bakr2011} for optical lattices. Thus the typical frequency in generating the spatial incoherence must be high. Additionally, the ultracold atom gas in an optical lattice is sensitive to noise and perturbations, so stability is also concerned. 

During the past decades, various ways of generating a spatial incoherent light based on the time averaging idea have been proposed and developed. A diffuser rotated by a motor is maybe the simplest way to destroy the spatial coherence \cite{Stangner:17}. And people can also use a galvo or an acousto optical deflector (AOD) to modulate the incident angle shining on a static diffuser to do so \cite{hudjf2o5h4eurcvau2dn88u0136bphoyt0rp16fi5w41k75n7r}. Another way is to use a multimode fiber (MMF) to turn the temporal incoherence into the spatial incoherence\cite{Fundamentals_of_Photonics}, which is the main focus of this work. Actually, the MMF can also be combined with other techniques (including diffusers, galvos, AODs, vibrators, and etc \cite{WOS:001010907100001,Mehta:12,Mao:23,Manni:12,Tong:17,7469822,Ha:09,photonics11030234}) to be applied in many situations. But for our specific purpose, we only discuss about the scenario of a static MMF associated with a LED source, for its several advantages: 

\textit{High modulation frequency}. For our purpose, the time averaging speed in the settings of a diffuser driven by a motor ($\sim 200$ Hz) or changing the light orientation by a galvo ($\sim 1$ kHz) are far insufficient, let alone the mechanical vibration noise they will introduce into the system. On the other hand, the effective  time averaging frequency of a MMF+LED method, characterized by the temporal coherence of the light source, can easily go up to around $1/\tau_c\sim 15$ THz for a LED with a $20$ nm line width. 

\textit{Stability}. Though associating the diffuser with an AOD may realize a modulation frequency about $100$ kHz, the output light beam will suffer a lateral shift due to the change of the diffraction angle in the AOD, which is also unwanted in our setting. In the MMF method, no movable component is included, and a high stability can be achieved. 

\textit{Practicality and cost}. A standard MMF is relatively cheap and easy to be obtained, making this approach more practical in experiments. And no fine adjustment is needed, which may be troublesome in the diffuser-AOD setting for instance. 

Properties of the MMF in the process of generating the spatial incoherence has already been studied these years, both experimentally and theoretically\cite{1074136,WOS:000338055900043,WOS:000500004100003,WOS:000486373100049,WOS:000454118600002,WOS:000446024300056}. We extent the former models in the context of DMD projection, trying to bridge between the previous extensive studies and the applications in the ultracold-atom experiments. A full-numerical model of the MMF, which is also compatible with other optical process including the free space propagation, is established, and part of the results is compared with our experimental data. We mainly investigate the intrinsic speckle contrast and the spatial correlation functions (or the perturbation robustness) of the MMFs in the projection process, which will be discussed in detail in the following sections. 

The article is organized as follows: we first study the influence of the fiber time delay ($t/\tau_c$) on the intrinsic speckle and the spatial correlation function, which is qualitatively consistent with former conclusions. Then we focus on how light propagation outside the MMF affects these two observables, showing that a magnifying 4f-system placed in between fiber endface and DMD is necessary. In the end, we compare MMFs with different radius choices, trying to give a guide of choosing the suitable parameters for the experiments using the DMD projection. 

\section{MMF projection settings and the numerical model}
\begin{figure}[htbp]
\centering
\includegraphics[width=0.8\linewidth]{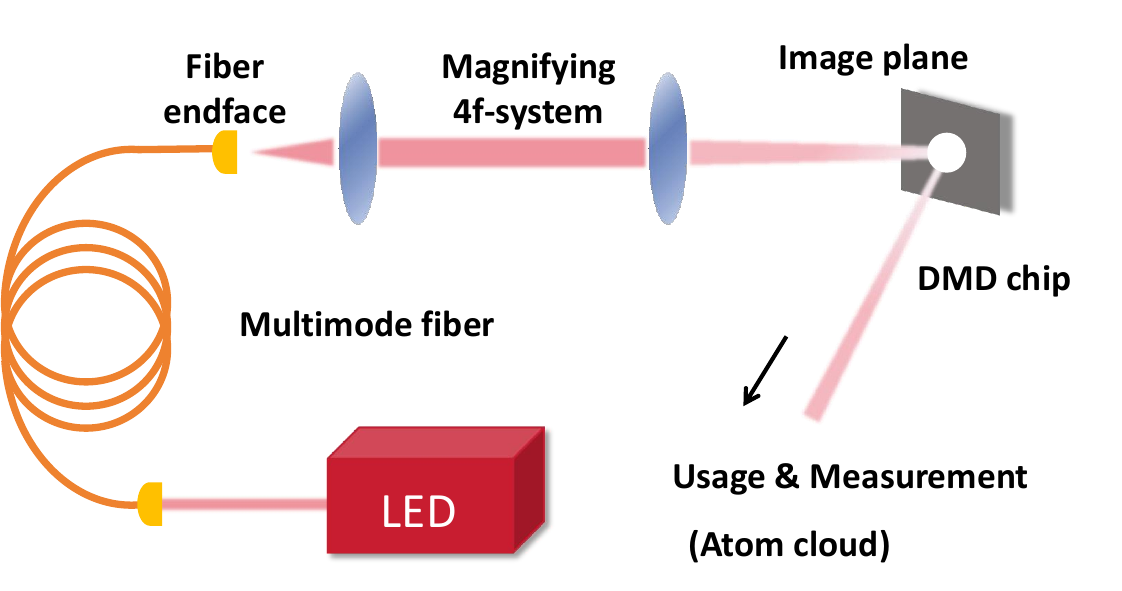}
\caption{An optical path design in the projection of a temporal-spatially incoherent light from a multimode fiber. The fiber endface is projected onto a DMD chip by a magnifying 4f-system. Arbitrary light potentials generated by the DMD are widely used in the ultracold-atom experiments. }
\label{fig1}
\end{figure}

We take the setting below to include the temporal-spatially incoherent light generated from a MMF into the standard DMD projection design in the ultracold-atom experiments, which is demonstrated in \figurename~\ref{fig1}. A temporally incoherent light from a LED source (EXS210017-01), is coupled into a round-core MMF (Thorlabs) with varying choices of core radius and length. The output of MMF (named ``endface'') is imaged by a magnifying 4f-system onto an image plane, where a DMD chip should be placed to project arbitrary patterns for usage. Note that the specific properties of the DMD are not our concentration, we only focus on the properties of the MMF light spot shone on the DMD chip (or the image plane). 

\begin{figure}[htbp]
\centering
\includegraphics[width=\linewidth]{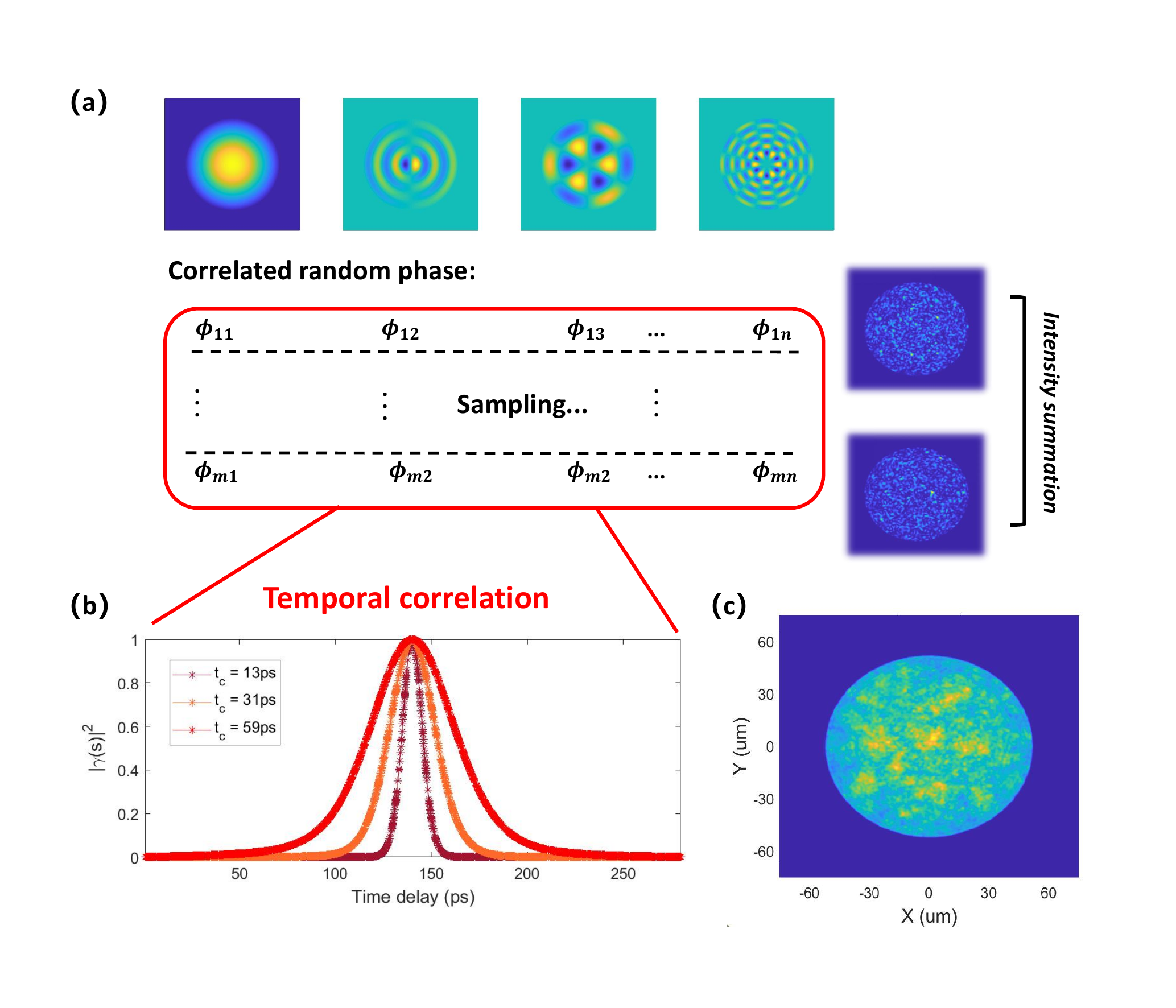}
\caption{Demonstration of our numerical model to simulate the producing of a spatially incoherent light in a MMF. (a) Every allowed mode in the fiber is given a phase which is correlated to that of the modes nearby. For every given set of phases, an intensity distribution can be calculated (named 'speckle images', see the right of (a)). And this process (named 'sampling') will be repeated for amounts of times. (b) The temporal correlation functions between the fiber modes with different coherence time of the light source. (c) A final and measurable intensity distribution at the fiber endface, obtained by summing all the 'speckle images' calculated in (a). }
\label{fig2}
\end{figure}

To closely study the properties of the light injected onto the DMD for use, we establish a full-numerical model to simulate the light propagation in the MMF, as shown in \figurename~\ref{fig2}. The intensity distribution at the MMF output (fiber endface) is the result of interference between all the radial modes, whose phases are not fixed with each other due to the fiber time delay and the finite temporal coherence length. An example of the fiber-supported radial modes are shown in \figurename~\ref{fig2}(a), by numerically solving the Helmholtz equation\cite{Fundamentals_of_Photonics}:
\begin{equation}
    \nabla^2 U + n^2k_0^2U = 0
\end{equation}
The complex amplitude function $U(\vec{r})$ can be decomposed into radial, angular, and axial components as: $U=u_m(r)e^{-j l\phi}e^{-j\beta_{lm} z}$, where $r$ and $\phi$ are radial coordinates in the fiber cross section, and $z$ is in the light propagation (axial) direction. $l=0,~\pm 1,~\pm 2,...$ represents the angular momentum, and $\beta_{lm}$ is the axial wave-number. We use $U_{lm}$ to label the radial mode with the angular momentum $l$ and the axial wave-number $\beta_{lm}$ as in \cite{Fundamentals_of_Photonics}. In the following, we mainly use a polarized model in analysis, where the polarization degeneracy of each $U_{lm}$ is neglected\cite{WOS:000454118600002}, while a vector version of our model is also taken into comparison in \figurename\ref{fig5}. The total number of the radial modes supported by a MMF is determined by its $NA$ and the size of core, and we have more than $2200$ modes for a standard MMF with $NA=0.22$ and core $D=105~\mu$m. 

The intensity distribution at the fiber output is generated by the following process: we first assign a phase to every radial mode in the fiber (here the phase correlation between the modes can be arbitrarily tuned, as will be explained below), and calculate their amplitude superposition, resulting in an intensity distribution with a speckle contrast near 1 (see \figurename~\ref{fig2}(a)). Then we repeat this process for amount of times (typically $\sim 1000$), and average over the intensity  of all these ``speckle pictures'' to get the ``real'' intensity distribution at the fiber endface, shown in \figurename~\ref{fig2}(c). Note that the validity of doing this comes from the fact that the camera exposure time ($\sim 40$ us) is much longer than the temporal coherence length (typically less than 300~ps), hence the image you get from a camera is actually the intensity summation of ``speckles'' at different time slides.

The key in conducting such a speckle generating process described above is about simulating the finite temporal coherence of the light source(or the phase correlation between the different modes). We do this by convolving a random phase list (corresponding to all the radial modes) with a gaussian kernel (representing a gaussian-shape spectrum of the light source) with a pre-determined width. The resulting temporal correlation functions $\gamma(\tau)$ between the assigned phases are shown in \figurename~\ref{fig2}(b), corresponding to different coherent times $\tau_c=\int_{-\infty}^{+\infty}|\gamma(\tau)|^2d\tau$. 

In the following, we calculate the intensity distributions at the endface for different fiber parameters. There are two observables extracted from the intensity distribution that can reflect the overall properties of the MMF: the intrinsic speckle contrast and the spatial correlation function \cite{WOS:000338055900043,WOS:000500004100003,WOS:000486373100049,WOS:000454118600002}. 

The intrinsic speckle contrast originates from both the facts that the number of the radial modes supported by MMF is finite and the temporal coherence length is non-zero, which means that there will still be some correlation between the phases of different modes, though it has been greatly suppressed by the MMF-caused time delay. The partially-coherent interference between all the radial modes gives this intrinsic speckle contrast, which is unwanted in the light projection, resulting in potential noise in DMD experiments for instance. 

We define this intrinsic speckle contrast as:
\begin{equation}
    C_s=\frac{std(I_{inter}/I_{incoh})}{\langle I_{inter}/I_{incoh}\rangle},
\end{equation}
here $I_{inter}$ is the calculated intensity distribution from our numerical model, and $I_{incoh}$ is that of an ideal incoherent superpositon of all the radial modes, which is nearly a smooth profile. The fluctuation of intensity is normalized by its mean value. 

The spatial correlation function is another quantity to estimate the spatial coherence of the light field coming out from the MMF, and is believed to be corresponding with the field's robustness against spatial perturbations (such as dusts and distortions in the optical path). It is defined as:
\begin{equation}
    \gamma(\vec{r_1},\vec{r_2})=\frac{U^*(\vec{r_1})U(\vec{r_2})}{\sqrt{|U(\vec{r_1})|^2\cdot |U(\vec{r_2})|^2}}
\end{equation}
Here $U(\vec{r})$ is the total amplitude of the light field expanded as $U(\vec{r})=\sum_{l,m}C_{lm}U_{lm}(\vec{r})$. 
%, where the axial coordinate $z$ has been neglected for simplicity. 
The coefficient $C_{lm}=|C_{lm}|e^{j\phi_{lm}}$ is a complex number containing information of both the occupation and the phase of each radial mode. In our model, we simply take $|C_{lm}|\equiv 1$ representing a uniform mode excitation. We choose $\langle|\gamma(s)|\rangle=\langle|\gamma(\vec{r},-\vec{r})|\rangle_{a,s/2}$ as an observable, where the $\langle\rangle_{a,s/2}$ stands for the angular average for all vector $\vec{r}$ with length s/2 on the fiber endface. In the following, we use both the residual coherence $\gamma_{rc}=\langle|\gamma(s\rightarrow\infty)|\rangle$ and the spatial coherence length $s_c=\int |\gamma(s)|^2ds$ to evaluate the spatial coherence, and find that in the case of DMD projection setup, $s_c$ is a better indicator for the light field robustness.  

Our model also simulates the light propagation in the free space which is included in the common DMD projection setting. The light propagation is governed by the Rayleigh-Sommerfeld integral formula\cite{Ersoy2006DiffractionFO}:
\begin{equation}
    U(x',y',z)=U(x,y,0)*h(x,y,z)
\end{equation}
in which the convolution kernel reads $h(x,y,z)=\frac{1}{j\lambda z}exp(jkz\sqrt{1+\frac{x^2+y^2}{z^2}})/(1+\frac{x^2+y^2}{z^2})$. 

We note that, compared to existing models\cite{WOS:000454118600002,WOS:000338055900043}, we provide here a full numerical method to investigate various properties of the partially temporal-spatially incoherent light generated from a MMF conveniently and directly. This model enables us not only to look into properties on the fiber endface, but also to study the propagation after the MMF, especially in the DMD projection setting.

\section{Results: Numerical and experimental}
\textbf{\textit{MMF length and temporal coherence}}. Our numerical model enables us to study the influence of both the fiber length and the temporal coherence length of the light source. Actually, the parameter really matters here is the ratio of the total time delay $t$ caused by the MMF to the temporal coherence length $\tau_c$. The MMF caused total time delay is expressed as\cite{Fundamentals_of_Photonics}:
\begin{equation}
    t=\frac{n}{c} (-1+\frac{1}{\sqrt{1-\frac{NA^2}{n^2}}})
    \label{eq:total-time-delay}
\end{equation}
where n is the refraction index of the fiber core, and c is the light speed in vacuum.

\begin{figure*}[htbp]
\centering
\includegraphics[width=0.75\linewidth]{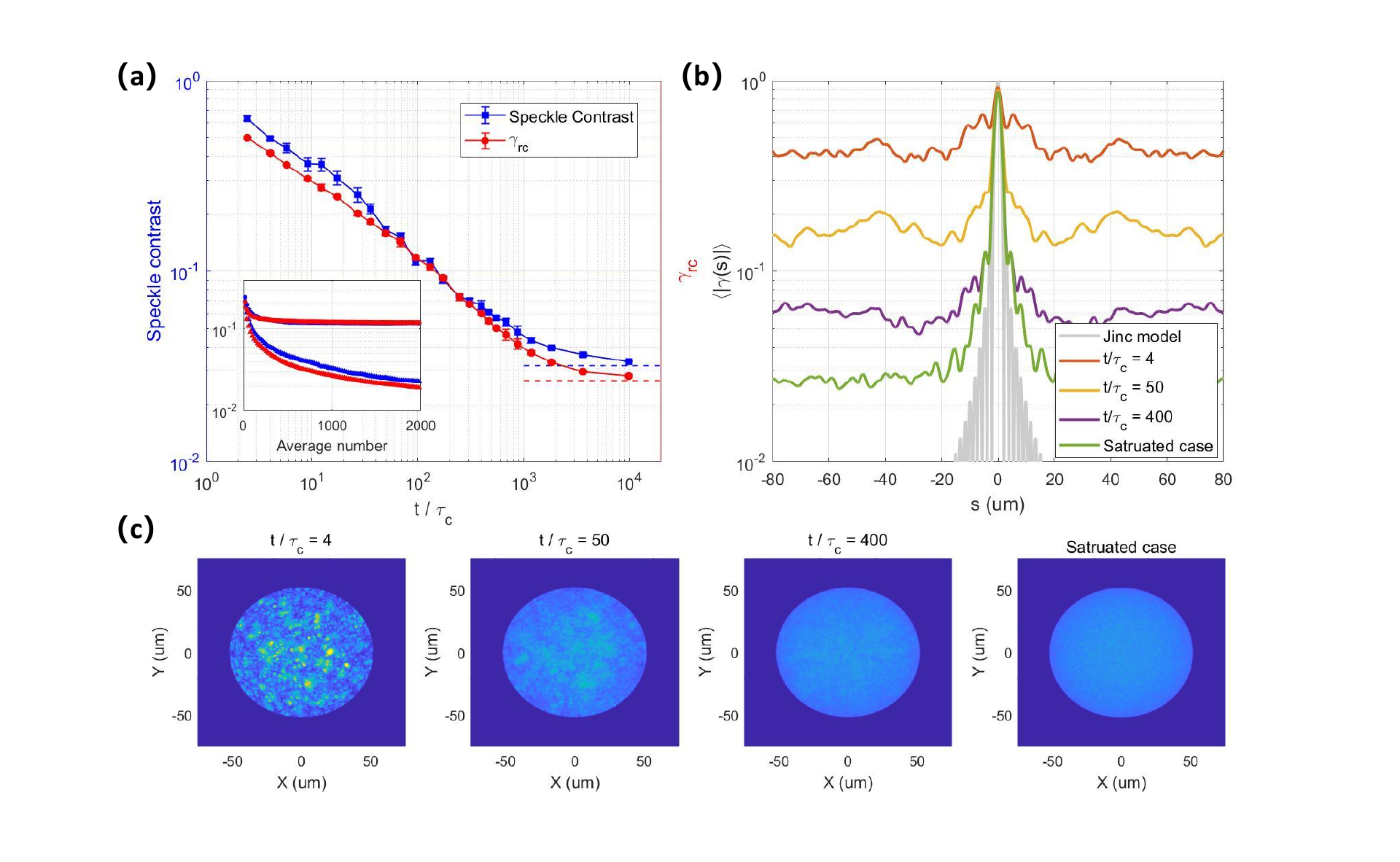}
\caption{Numerical results of the intrinsic speckle contrast and the spatial coherence at the fiber endface. Fiber parameters: $NA=0.22$, $D=105$ $\mu$m, $L=5$ m, with a total modal time delay $t=280$ ps as defined in Eq.~\eqref{eq:total-time-delay}. (a) The intrinsic speckle contrast and the residual coherence as functions of $t/\tau_c$. The blue (red) dash lines correspond to the saturation value of the intrinsic speckle contrast (the residual coherence). The errorbar represents the fluctuations in the numerical simulation. Inset: influence of numerical averaging (sampling) times on the results, the upper two lines correspond to $t/\tau_c= 100$, while the lower two lines correspond to the case near saturation levels.
Typically we choose 1000 average times in this work. (b) The spatial correlation functions at the fiber endface. The original 2D correlation function is averaged for both angular direction and different repeats. (c) Demonstrations of the intensity distribution at the fiber endface under different $t/\tau_c$ choices. }
\label{fig3}
\end{figure*}

\figurename~\ref{fig3} shows the dependence of both the intrinsic speckle contrast and the spatial correlation function on this dimensionless parameter $t/\tau_c$. In \figurename~\ref{fig3}, the MMF parameters (NA, core radius, and fiber length) are fixed and we only change the temporal coherence $\tau_c$ of the light source (see \figurename~\ref{fig2}(b)). \figurename~\ref{fig3} reveals that both the intrinsic speckle contrast and the residual coherence decrease monotonously with $t/\tau_c$, as expected intuitively. The corresponding spatially correlation functions and endface intensity distributions are demonstrated in \figurename~\ref{fig3}(b) and (c). Results above are in good agreement with former models\cite{WOS:000454118600002}. We note here that, the saturation behavior at the large $t/\tau_c$ side is a numerical effect, and the saturation limit is limited by the finite sampling time in our model, as shown in the inset of \figurename~\ref{fig3}(a). 

\textbf{\textit{Light propagation after the MMF}}. The typical core radius of a MMF varies from tens to hundreds of micrometers, far smaller than the DMD chip to be illuminated, which is in scale of $\sim 1$ cm. Here we points out the necessity of a magnifying 4f-system over the free divergence of the light from the MMF endface. 

\begin{figure*}[htbp]
\centering
\includegraphics[width=0.75\linewidth]{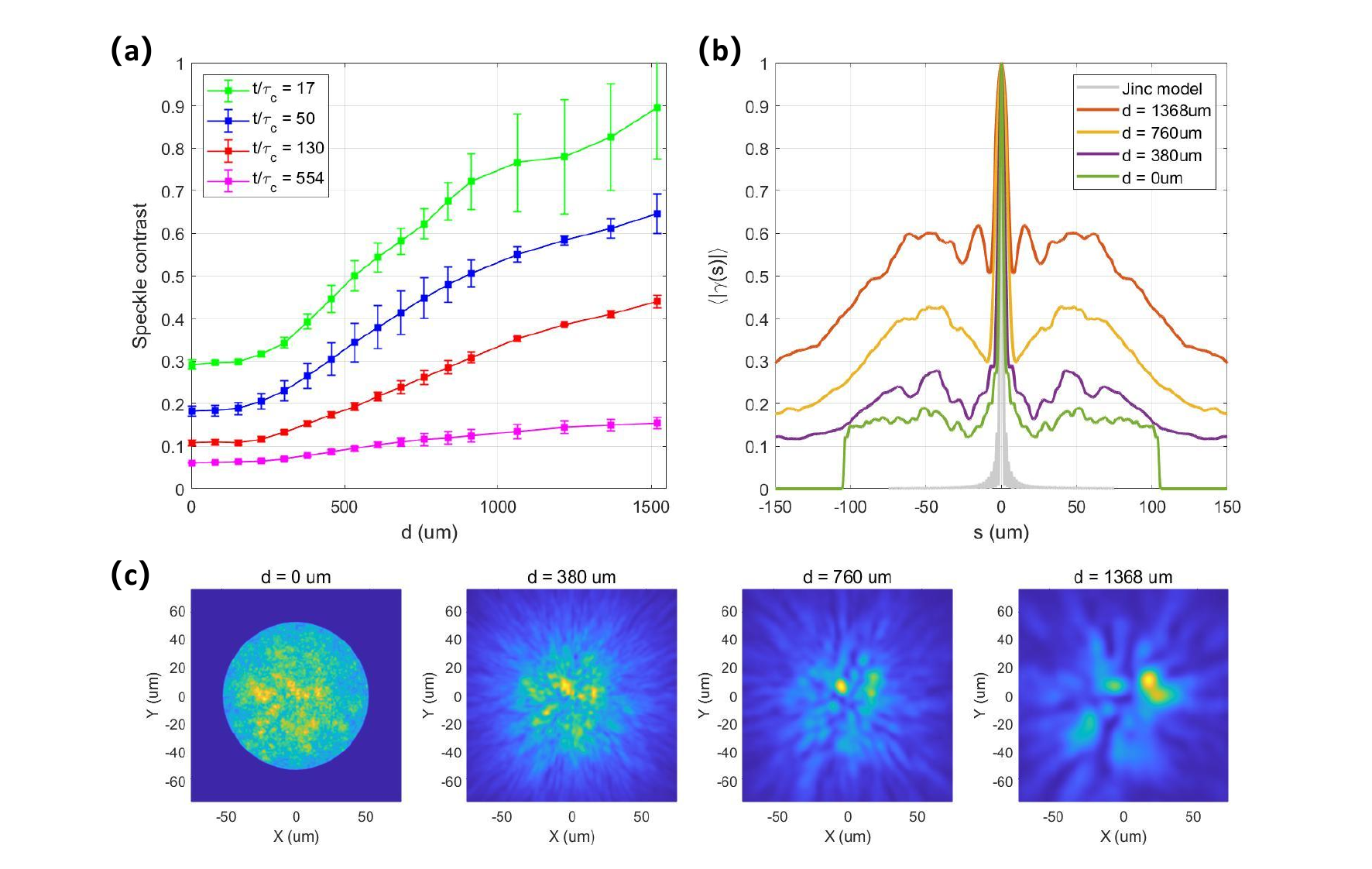}
\caption{The numerical results of the intrinsic speckle contrasts and the spatial correlation functions after propagation directly from the fiber endface. Fiber parameters: $NA=0.22$, $D=105$ $\mu$m, $L=5$ m. (a) The dependence of the intrinsic speckle contrast on the propagation distance $d$ at different $t/\tau_c$. The errorbar represents the fluctuations in the numerical simulation. (b)(c) The spatial correlation functions and intensity distributions during the light propagation, with $t/\tau_c= 50$. }
\label{fig4}
\end{figure*}

\figurename~\ref{fig4} shows the evolution of both the intrinsic speckle contrast and the spatial correlation function in the free light propagation out of a MMF endface.
\figurename~\ref{fig4}(a) clearly shows that, after only $\sim 200$ $\mu$m out of the MMF endface, the intrinsic speckle contrast shows a significant rise, which corresponds to the change of spatial correlation function in \figurename~\ref{fig4}(b). The response of the intrinsic speckle contrast becomes milder when $t/\tau_c \rightarrow 0$, though this effect is still unwanted. 

\begin{figure*}[htbp]
\centering
\includegraphics[width=0.75\linewidth]{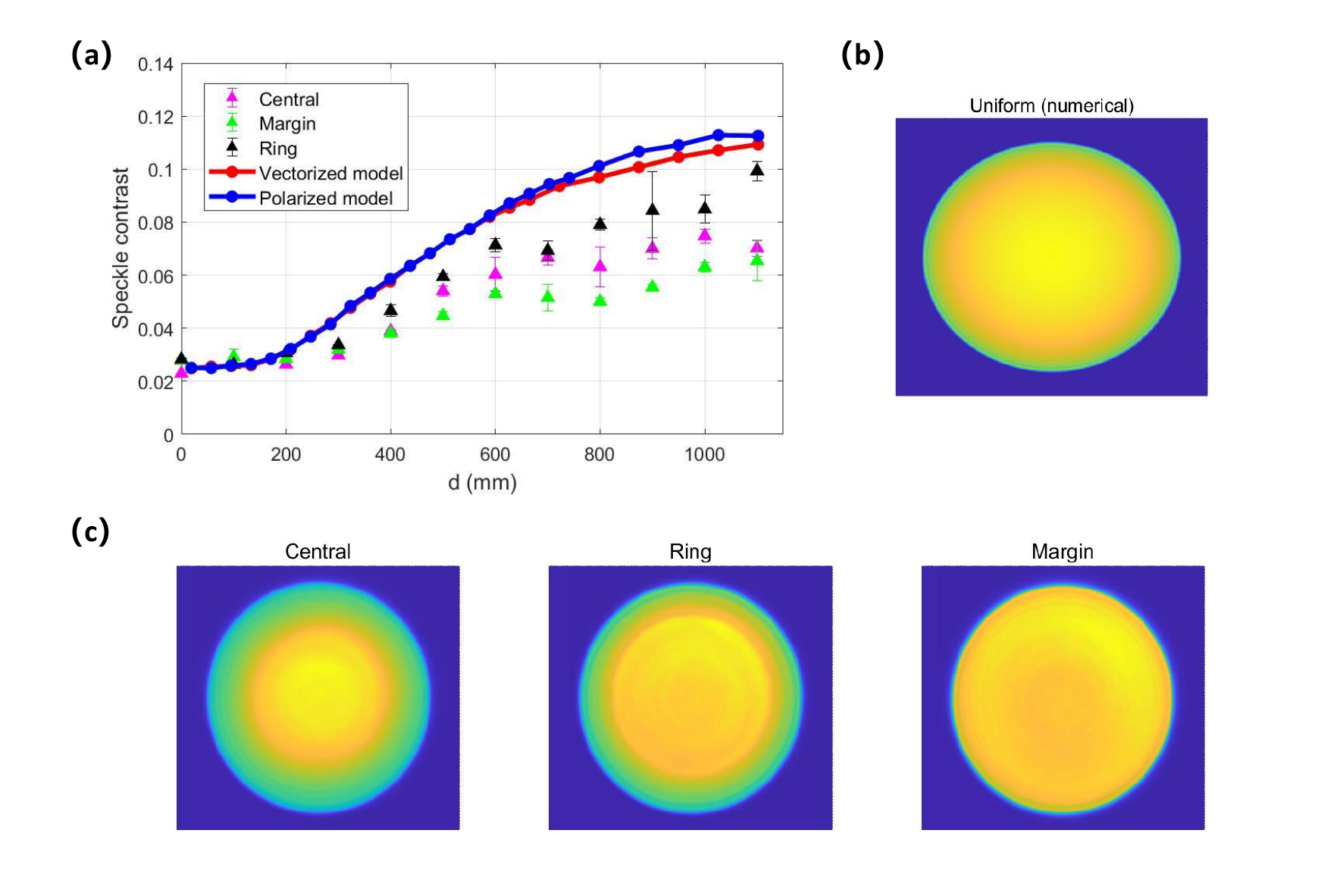}
\caption{The intrinsic speckle contrast after the magnifying 4-f system. Fiber paramters: $NA=0.22$, $D=105$ $\mu$m, $L=1$ m. The light source has a temporal coherent time about 66 fs, and the 4f-system magnification ratio is $M=26$. (a) The experimental and numerical results of the intrinsic speckle contrast as a function of the defocus distance $d$, which is the distance between the measured plane (where the DMD should be placed) and the focus plane of the magnifying 4-f system. (b) The uniform mode excitation in the numerical simulation. (c) Three different mode excitation patterns in the experiment. }
\label{fig5}
\end{figure*}

Instead, the situation will be improved when we add a magnifying 4f-system to image the MMF endface directly onto the DMD chip (``image plane'' in \figurename~\ref{fig1}), as shown in \figurename~\ref{fig5}. We measure the intrinsic speckle contrast evolution after the magnifying 4f-system as the function of defocus distance relative to the focal plane. In the experiment, we use a MMF with a radius $105$ $\mu$m and a length $1$ m (Thorlabs M15L01) to compare with the numerical results, here a shorter fiber length will enlarge the measured speckle signal. Our light source (EXS210017-01) has a temporal coherence length about $66$~fs, and the magnification ratio is set about $26$ for experimental convenience. 

In the experiment, we test three different mode excitation conditions (central, ring and marginal excitation) by adjusting the in-coupling light angle at the input of the MMF, and we compare their results with numerical simulations in \figurename~\ref{fig5}(a). In the numerical simulation, we consider both the polarized and vectorized model, under a uniform mode excitation condition. All simulated speckle contrasts have been subtracted from a numerical saturation background value (horizontal dash line \figurename~\ref{fig3}(a)) to match their experimental counterparts.

Qualitatively, the simulation is in consistence with experimental results. At the DMD chip (or the image plane), the intrinsic speckle contrast stays the same as at the fiber endface. When measured at a defocus plane ($d$ after the original image plane), the intrinsic speckle contrast show an upward trend with increasing $d$. The quantitative difference is mainly due to the mode excitation difference comparing to an ideal uniform distribution in the simulation.

The necessity of this magnifying 4f-system mainly shows in two aspects. On the one hand, compared to a direct illumination of DMD from the fiber endface, we avoid the unnecessary enhancement of the intrinsic speckle contrast due to the light propagation when a magnifying 4-f system is applied. On the other hand, we note from \figurename~\ref{fig5}(a) that, the intrinsic speckle contrast stays relatively insensitve to the defocus distance below $d=200$ mm. Actually, this insensitive region is also seen in the free propagation case, as shown in \figurename~\ref{fig4}(a), for a range about 200~$\mu$m. The magnifying 4-f system has enlarged the insensitive distance for almost three orders. The existence of this large insensitive region indicates that the intrinsic speckle magnified on the DMD chip can be quite robust against a small defocus in the system, and no subtle optical adjustment is needed. 

Besides, \figurename~\ref{fig5}(a) reveals that intrinsic speckle contrast's dependence on the defocus distance is influenced by the actual mode excitation condition, which also explains the deviation between experimental and numerical results. Among all the experimental data, the ring excitation one seems to be the closest to the numerical simulation, corresponding to the fact that its mode excitation pattern looks the most similar to the numerical uniform mode excitation, as shown in \figurename~\ref{fig5}(b) and (c). 

\textbf{\textit{MMF radius and robustness against perturbation}}. We now study how the performance of a MMF is influenced by its core radius. A MMF with a larger core radius will support more radial modes in the scale of $R^2$\cite{Fundamentals_of_Photonics}. In the meantime, the total time delay through the MMF is only determined by its NA, which is fixed at $0.22$ in our discussion. Thus a larger core radius results in a denser mode distribution along the time axis. Intuitively, this will influence the dependence of the intrinsic speckle contrast on the parameter $t/\tau_c$ (see \figurename~\ref{fig3}), resulting in a later saturation. On the other hand, the existence of more radial modes means that the superposition of the light field will become more chaotic, and may lead to a better robustness against perturbations in the real optical paths. In the following, we compare the MMFs with various core radii in these two aspects. 

\begin{figure*}[htbp]
\centering
\includegraphics[width=0.75\linewidth]{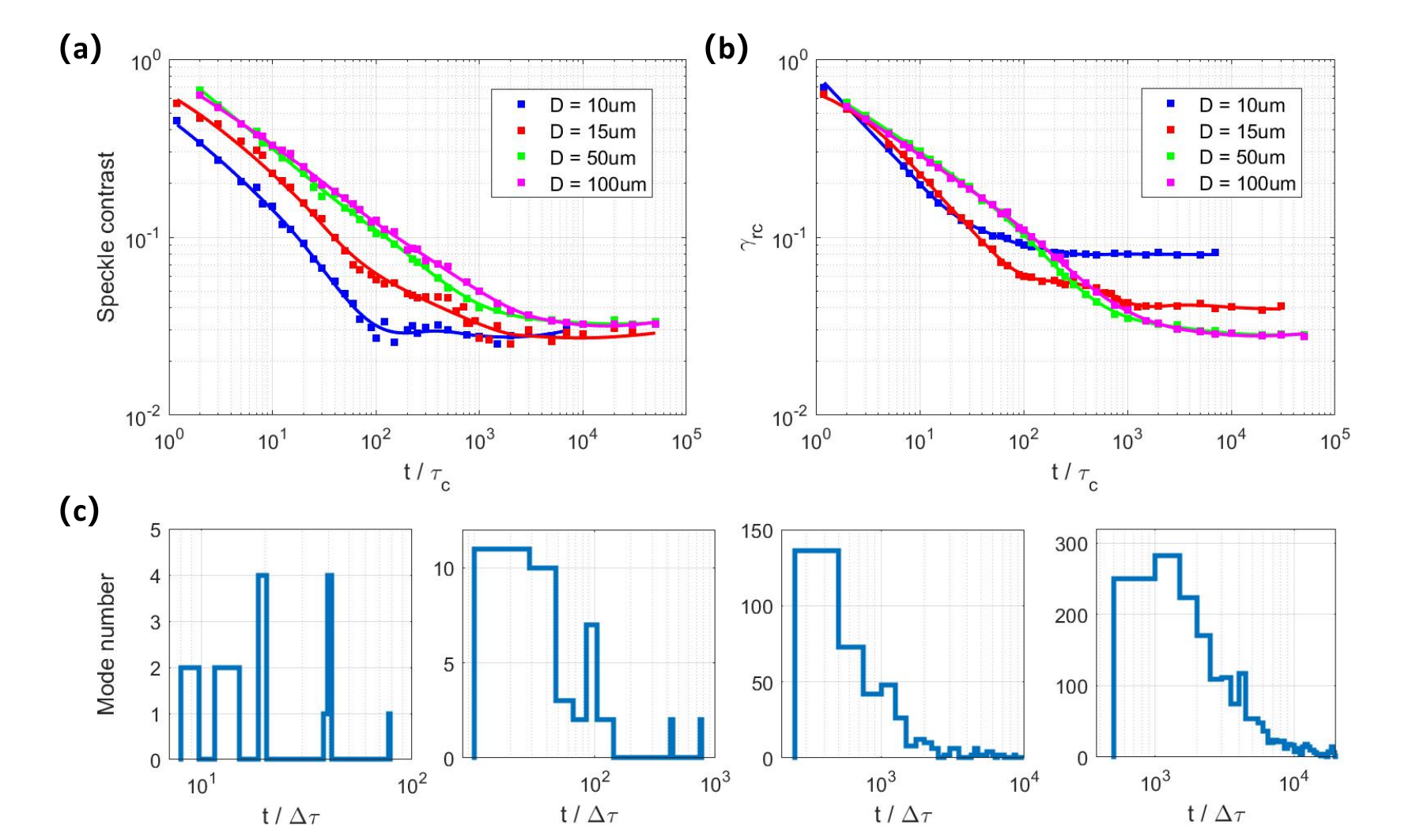}
\caption{The saturation behavior for MMFs with various core radii of (a) the intrinsic speckle contrast and (b) the residual spatial coherence given by the numerical model. Fiber with a smaller radius shows a earlier saturation and a higher saturated residual coherence. Solid lines are guide lines. (c) The distribution of the time delay between all the radial modes after the MMF. $\Delta \tau$ stands for the temporal separation between two adjacent modes, and t is the MMF's total time delay given by Eq.~\eqref{eq:total-time-delay}.} 
\label{fig6}
\end{figure*}

The saturation behaviour of both the intrinsic speckle contrast and the residual coherence for different fiber core diameters (10, 15, 50, and 100 $\mu$m) are shown in \figurename~\ref{fig6}(a) and (b), respectively. \figurename~\ref{fig6}(c) plots out the distributions of the differences between time delays of two adjacent radial modes (labeled as $\Delta \tau$) in MMFs with the corresponding radii. 

For the intrinsic speckle contrast, it's seen that MMF with a larger core radius shows a slower saturation with the increasing $t/\tau_c$, and the different MMFs saturate at the same value restricted by numerical sampling times (see the inset of \figurename~\ref{fig3}(a)). We note here that the plateaus (like that in $D=15$ $\mu$m curve) and turning points of the curves in \figurename~\ref{fig6}(a) accurately correspond to the distribution peaks in \figurename~\ref{fig6}(c). It can be explained by an intuitive picture as follows: Firstly, it's known that only when the temporal separation between two adjacent modes $\Delta \tau$ becomes comparable with the light source's temporal coherent length, the interference becomes incoherent. Secondly, \figurename~\ref{fig6}(c) shows that the temporal separation $\Delta \tau$ actually has a distribution thus modes become uncorrelated gradually, especially when the fiber core radius is large. Therefore there is a clear turning point at $t/\tau_c\sim 100$ in the $D=10~\mu m$ curve, and two turning points along with a plateau in between emerge in the $D=15~\mu m$ curve, while the curves of $D=50~\mu m$ and $100\mu m$ become smooth and no obvious turning point is shown. 

This slow down due to an increasing core radius can also be seen in the change of the residual coherence in \figurename~\ref{fig6}(b). Note that a MMF with a larger core radius shows a smaller saturated value of $\gamma_{rc}$, which may indicate its less spatial coherence and better robustness against perturbations. Here the curves of $D=50$ $\mu$m and $D=100$ $\mu$m coincide at the large $t/\tau_c$ limit, which we also attribute to the finite numerical sampling times. Besides, a curve 'crossing' appears when $t/\tau_c$ is in the range of $10\sim 100$. This phenomenon may reveal the competition of two mechanisms in the fiber: comparing MMFs with different radii, when $t/\tau_c$ is not large enough, different radial modes interfere more coherently, thus the existence of denser modes results in a greater spatial correlation. However, when $t/\tau_c$ is chosen relatively large, the phase correlation between the radial modes will be significantly suppressed, and the interference becomes  more incoherent, therefore the denser modes make the light field more stochastic and reduce the spatial correlation level. 

Results in \figurename~\ref{fig6} give us some instructions about choosing the suitable MMFs in the experiments. On the one hand, in the large $t/\tau_c$ limit, MMFs with larger radii become more spatially incoherent and may endure more severer perturbations. On the other hand, if choosing a large radius MMF but with a relatively small $t/\tau_c$ (fiber too short or linewidth of the source too narrow), the intrinsic speckle will be worse, as well as the residual coherence. 

\begin{figure*}[htbp]
\centering
\includegraphics[width=0.75\linewidth]{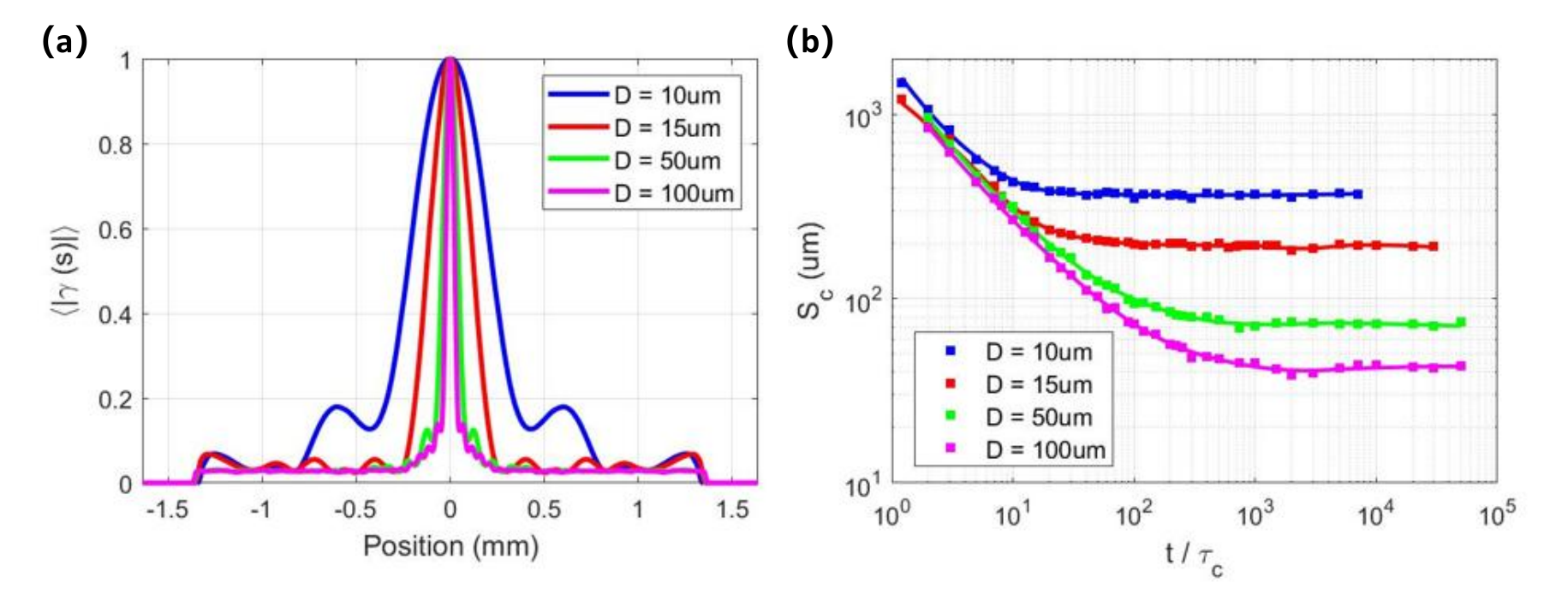}
\caption{(a) The spatial correlation functions of MMFs at the image plane (the DMD chip) after the magnifying 4f-system given by the numerical model. All MMFs with various radii are imaged into the same size by adjusting the magnification ratio. (b) The spatial coherent length $s_c$ as a function of $t/\tau_c$ for MMFs with different radii. }
\label{fig7}
\end{figure*}

A further question arises here that, whether the residual coherence is the only important and sufficient indicator to summarize the spatial correlation function of a MMF? In \figurename~\ref{fig7}, we point out that on the image plane of the magnifying 4f-system (where we plane to place the DMD chip), the spatial coherence length \cite{hudjf2o5h4eurcvau2dn88u0136bphoyt0rp16fi5w41k75n7r}:
\begin{equation}
    s_c=\int_{-\infty}^{+\infty} |\gamma(s)|^2 ds
\end{equation}
may be another choice. In \figurename~\ref{fig7}, different magnification ratios are chosen to image the MMF endfaces with various core radii into the same shape on the DMD chip (or the image plane, where we do the measurement). We discover that in this sense, the central peak of the correlation function actually plays a more dominant role than we thought, from both the $\langle|\gamma(s)|\rangle$ shapes in \figurename~\ref{fig7}(a) and the split of the curves of $D=50$ $\mu$m and $D=100$ $\mu$m at the large $t/\tau_c$ limit in \figurename~\ref{fig7}(b). Another fact noted here is that, there are no crossings between the MMFs with different core radii, unlike in \figurename~\ref{fig6}(b) ---- MMF with a larger core radius always seems to be more spatially incoherent from the view of $s_c$.  

\begin{figure*}[htbp]
\centering
\includegraphics[width=0.75\linewidth]{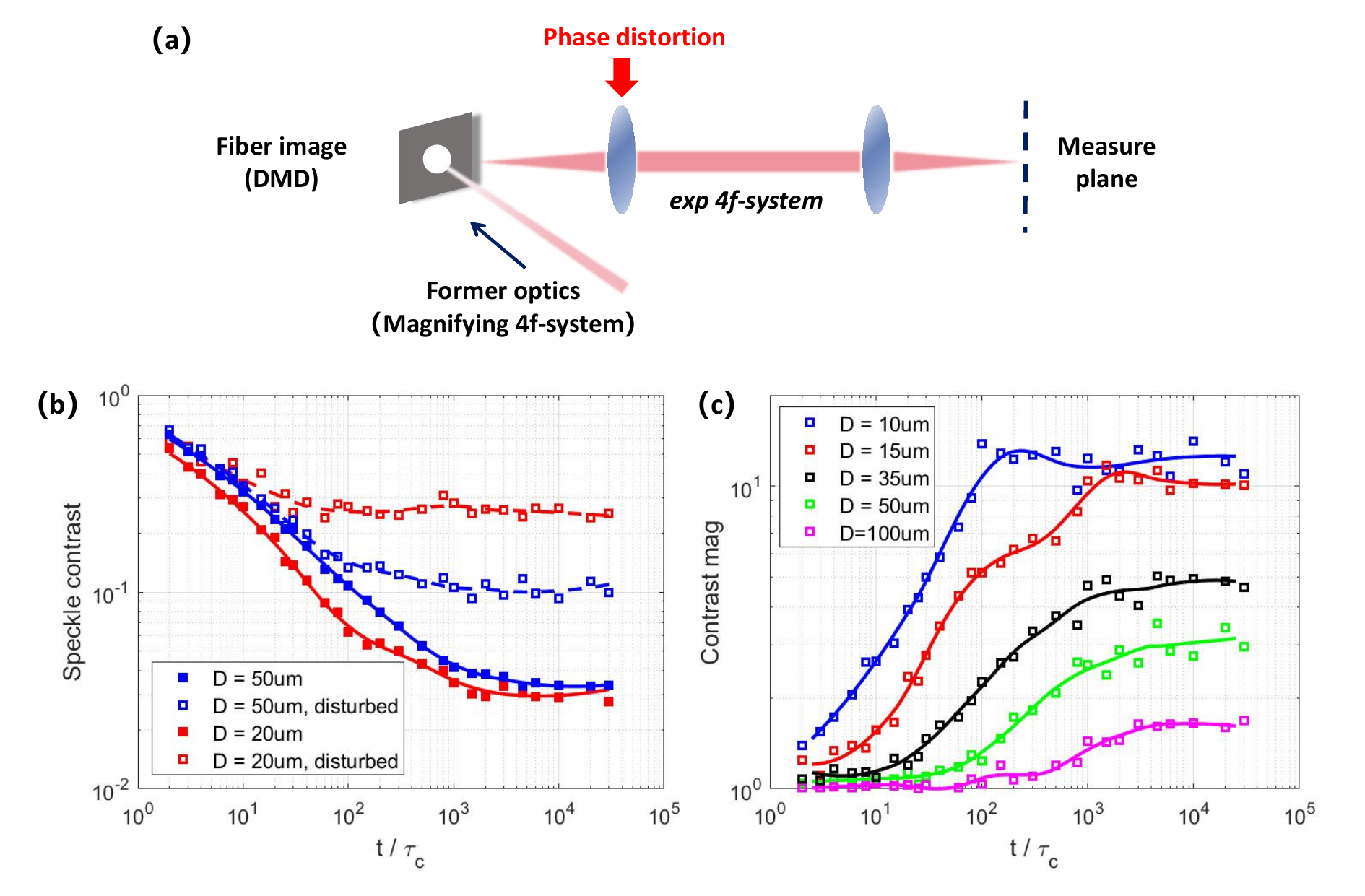}
\caption{Response of the MMFs with different radii in an exp 4f-system where artificial distortions are imposed, given by the numerical model. (a) All MMFs are imaged onto the DMD chip into the same shape (as in \figurename~\ref{fig7}). After that, an exp 4f-system with $f=140$ mm and magnification ratio $1$ images the magnified light spot onto a measure plane. A phase distortion with an amplitude of $0.2\pi$ and a typical scale  of $300$ $\mu$m is imposed on the first lens. (b) Demonstration of the $C_s-t/\tau_c$ curves for two MMFs. Solid squares correspond to ideal imaging, while hollow squares for disturbed imaging. (c) The contrast magnification $C_{s,dist}/C_{s}$ for MMFs with various radii. }
\label{fig8}
\end{figure*}

We then try to explicitly determine which way of evaluating the spatial coherence, the residual coherence $\gamma_{rc}$ or the spatial coherence length $s_c$, is closer to the truth. We do this by imposing an artificial phase distortion on the MMF light field and measuring its response. The setting is like this: after the MMF endface is magnified onto a DMD chip, another 4f-system without further magnification (noted as the 'exp 4f-system' and we choose $f=140$ mm for numerical convenience) images the light spot at DMD onto a new 'measure plane', as shown in \figurename~\ref{fig8}(a). In the ideal case, no properties of the light spot will be changed by this perfect 4f-system. 

To test the light field's response (or robustness) to the external perturbations in the optical path, we artificially add a phase distortion at the first lens of the ‘exp 4f-system’. We choose an amplitude of distortion around $0.2\pi$ (or $\lambda/10$), and a typical length $300$ $\mu$m to simulate the dusts or roughness on the optical surface. We compare the disturbed speckle contrast at the measure plane with that of the undisturbed one, and use the ratio $C_{s,dist}/C_s$ to quantitatively evaluate the robustness of the MMFs. Note that in the 'exp 4f-system', for all the MMFs with different radii, the light spots are actually in the same size (due to the magnifying 4f-system described in \figurename~\ref{fig7}), so the difference in their performances against the distortion will originate only in their internal spatial correlation properties.

\figurename~\ref{fig8}(b) demonstrates the response of $D=20$ $\mu$m and $D=50$ $\mu$m fibers. It can be seen that the whole curve is raised when distortion is brought in, and the fiber with a smaller core radius saturates at a higher level. \figurename~\ref{fig8}(c) gives more data about the contrast magnification ratio $C_{s,dist}/C_s$ as a function of $t/\tau_c$.  We see that MMF with a larger core radius shows a minor response and eventually saturates at a smaller ratio. Comparing this result with \figurename~\ref{fig6}(b) and \figurename~\ref{fig7}(b), the spatial coherence length $s_c$ seems to serve as a better indicator for the light robustness , especially in the DMD projection settings, than the residual coherence $\gamma_{rc}$ which has been extensively studied before.

\section{Conclusions and discussions}
This work establishes a full-numerical model to investigate the various properties of the MMF in the DMD projection setting, which is widely used in the ultracold-atom experiments. 
This numerical model gets ride of many sophisticated analytic derivations concerning the speckle contrast and the correlation functions, and it can also be easily applied into other optical processes such as the light propagation in free space or with random phase distortions. For some basic properties of the MMF such as the role of $t/\tau_c$ in \figurename~\ref{fig3}, this model has led to similar conclusions with former works. Unlike previous researches, we pay more attention on two aspects: 
The first is how the intrinsic speckle contrast and the spatial correlation function are affected by the free space propagation of the light out of the MMF, and we point out the necessity of a 4f-based imaging of the fiber endface in the projection setting. The second is that, for the spatial correlation functions of MMFs with various radii, we show that in our DMD-projection setting, the spatial coherence length $s_c$ is a more reliable parameter than the often studied residual coherence $\gamma_{rc}$. The results in this work provide a guide about how to set up incoherent light source for the ultracold-atom experiments.

\section{Acknowledgments}
The authors thank Shuyao Mei, Ye Tian, Yuqing Wang and Tao Zhang for their helpful discussions and suggestions. 

%%%%%%%%%%%%%%%%%%%%%%%%%%%%%%%%%%%%%%%%%%%%%
\newpage
\bibliography{main}

\end{document}